\newcommand{\diracslash}[1]{#1\llap{/\kern2pt}}

\def\bearr{\begin{eqnarray}}

\def\eearr{\end{eqnarray}}

\newcommand{\be}{\begin{equation}}

\newcommand{\ee}{\end{equation}}

\newcommand{\bea}{\begin{eqnarray}}

\newcommand{\eea}{\end{eqnarray}}

\newcommand{\ba}[1]{\begin{array}{#1}}

\newcommand{\ea}{\end{array}}

\newcommand{\eqrf}[1]{Eq.\ (\ref{#1})}

\documentclass[preprint,secnumarabic,floats,amssymb,bibnotes,nofootinbib,tightenlines]{revtex4}

\usepackage{graphicx}
\usepackage{fixltx2e}
\usepackage{amsmath}
\usepackage{amssymb}
\usepackage{tabularx}
\usepackage{romannum}

\usepackage{latexsym}

\usepackage{epsfig}

\usepackage{subfigure}
\usepackage{hyperref}
\usepackage[outdir=./]{epstopdf}
\usepackage{placeins} 

\begin{document}
\title{Power law $\alpha$-Starobinsky inflation}
\author{Saisandri Saini and Akhilesh Nautiyal}
\affiliation{Department of Physics, Malaviya National Institute of Technology Jaipur,
JLN  Marg, Jaipur-302017, India}

\begin{abstract}
In this work we consider a generalization of Starobinsky inflation obtained by combining power law ($R^\beta$),  
and $\alpha$-Starobinsky inflation ($E$-model).  The  Einstein frame potential for this model is that of power law Starobinsky
inflation modified by a parameter $\alpha$ in the exponential. After computing power spectra for scalar and tensor perturbations
numerically, we perform MCMC analysis to put constraints on the potential parameters $\alpha$, $\beta$ and $M$, and the number
of e-foldings $N_{pivot}$ during inflation, using Planck-2018, BICEP/Keck (BK18), DES and BAO observations.
We find $\log_{10}\alpha= 0.37^{+0.82}_{-0.85}$, $\beta = 1.969^{+0.020}_{-0.023}$, 
$M=\left(3.54^{+2.62}_{-1.73}\right)\times 10^{-5}$ and $N_{pivot} = 47\pm{10}$. With these mean values of the potential parameters $\alpha$ and $\beta$, and varying
$N_{pivot}$ between $40$ to $55$, we also find that
the $r-n_s$ predictions of our model
lie well within the $1\sigma$ bounds of joint constraints from combined
analysis of ACT, Planck-2018, BICEP and BAO
observations. We compute the Bayesian evidences for our proposed
model, power law Starobinsky inflation, $\alpha$-Starobinsky inflation and Starobinsky inflation. Considering the Starobinsky 
model as the base model, we calculate the Bayes factor and find that our proposed model is mildly favored by the CMB and LSS 
observations.
\end{abstract}
\maketitle
\section{INTRODUCTION}
Inflation \cite{Guth:1980zm} refers to a period of rapid expansion in the early Universe before big-bang nucleosynthesis, 
proposed  to address several challenges of the standard big bang cosmology such   as the horizon problem,  
flatness problem, and the puzzling absence of magnetic monopoles  \cite{Linde:1981mu,Starobinsky:1980te}. 
Inflation not only provides an explanation for these issues but also turns them into natural outcomes of the theory. 
It is achieved by a scalar field, named as inflaton, whose potential energy dominates the energy density of the universe
causing quasiexponantial expansion of the universe for a very short period of time. 
Inflation produces density perturbations due to quantum 
fluctuations in the scalar field that are coupled to fluctuations in the metric, and tensor perturbations (primordial 
gravitational waves), which are generated due to the quantum fluctuations in the spacetime geometry. 
 These quantum fluctuations  serve as the seeds for  the large scale structure  of the Universe 
and cosmic microwave background anisotropy \cite{Mukhanov:1981xt,Starobinsky:1982ee,Guth:1985ya}. The predictions of  inflation 
i.e., adiabatic, nearly-scale invariant and Gaussian perturbations  are firmly verified with the various CMB and LSS 
observations such as COBE \cite{COBE:1992syq}, WMAP \cite{WMAP:2010qai}, 
Planck \cite{Planck:2015sxf, Planck:2018jri} and BICEP-Keck array 
\cite{BICEP:2021xfz}.

While inflation is generally seen as a successful idea, we still have not found a unique model of inflation that fits well 
within observational limits which have become more precise overtime. Since its inception, many different inflationary models have 
been proposed and examined \cite{Martin:2006rs},\cite{Martin:2013tda} with observational data. 
Analysis of recent Planck 2018 results \cite{Planck:2018jri} shows that the simplest classic inflaton potentials, 
$\frac{1}{2}m^2\phi^2$ and $\lambda\phi^4$, are  strongly ruled out. As proposed by Starobinsky \cite{Starobinsky:1980te}, 
inflation can also be achieved without scalar field by adding $\frac{1}{M}R^2$ term, $R$ being the Ricci scalar, in the 
Einstein-Hilbert action. Named as Starobinsky inflation, this model is in remarkable agreement with  
Planck-2018 \cite{Planck:2018jri} and BICEP/Keck \cite{BICEP:2021xfz,Tristram:2021tvh} observations as it predicts lower values 
for tensor-to-scalar ratio $r$.  
Originally described in the Jordan frame, Starobinsky inflation can be reinterpreted in the Einstein frame through a conformal 
transformation,  where it looks like a single field inflationary model with a potential that flattens out at large field values 
ensuring a slow and sustained period of inflation. Interestingly this potential in the Einstein frame can be easily derived 
from no-scale supergravity with a non-compact $SU(2,1)/SU(2)\times U(1)$ symmetry \cite{Ellis:2013xoa}, where we have a modulus 
field fixed by the other dynamics and the inflaton field is a part of the chiral superfield with a minimal Wess-Zumino 
superpotential. This enables incorporating inflation with various other aspects of particle physics phenomenology 
\cite{Ellis:2016ipm,Ellis:2017jcp,Ellis:2019opr}. However, analysis of recently reported Atacama Cosmology Telescope data
\cite{ACT:2025fju}, in combination with
Planck 2018, BAO and DESY Y1 data, indicates that the Starobinsky model is disfavored at the $2\sigma$ level
\cite{ACT:2025tim}.

Several generalizations of  Starobinsky inflation have been proposed, such as, power law Starobinsky inflation
having $R^\beta$ term in the Einstein-Hilbert action    
\cite{Schmidt:1989zz,Maeda:1988ab,Muller:1989rp,Gottlober:1992rg,DeFelice:2010aj,Nojiri:2010wj,Nojiri:2017ncd}, which is based
on higher-order metric theories of gravity, and $\alpha$-Starobinsky inflation ($E$-model) based on supergravity 
inspired deformation of the inflaton potential in the Einstein frame \cite{Ellis:2013nxa,Ferrara:2013rsa,Kallosh:2013yoa}.  
Although inspired by 
higher-order metric theories of gravity, the potential for the power law Starobinsky inflation can also be 
derived from no-scale supergravity \cite{Chakravarty:2014yda}. 
 Power law Starobinsky inflation gained traction in 2014 when BICEP2 reported a large value of tensor-to-scalar 
ratio $r=0.2^{+0.07}_{-0.05}$  \cite{BICEP2:2014owc}, as it was shown that this model generates a large $r$ compared to 
Starobinsky inflation for $\beta$ slightly less than $2$ \cite{Codello:2014sua,Costa:2014lta,Martin:2014lra}. 
The primordial origin of BICEP2 signal of $B$-mode polarization was ruled out later and it was found that it is due to the 
foreground dust emission \cite{Planck:2014dmk}. Observational constraints on $R^\beta$ inflation have been studied in 
\cite{Chakravarty:2014yda,Motohashi:2014tra,Odintsov:2022bpg,Meza:2021xuq,Saini:2023uqc} and it is found that the current
data allows a slight deviation from $\beta=2$ (see \cite{Saini:2023uqc} for a detailed statistical analysis with Planck-2018 
and BICEP3/Keck array data). 
In $\alpha$-Starobinsky inflation the scalaron potential in the Einstein frame contains a parameter $\alpha$ in the 
exponential that modifies the predictions for $n_s$ and $r$.  
The potential for this model can be obtained by  generalizing the coefficient of the logarithm of the volume modulus field in 
the K\"ahler potential in $SU(2,1)/SU(2)\times U(1)$ no-scale supergravity with a suitable choice of the superpotential 
having both the volume modulus field and the chiral superfield.   
The parameter $\alpha$ is these models is related to the geometry of the K\"ahler manifold.
The observational constraints on $\alpha$-Starobinsky inflation from Planck and 
BICEP/Keck array data are studied in 
\cite{Planck:2015sxf,Planck:2018jri,Ueno:2016dim,Ellis:2020xmk,Ellis:2021kad,ElBourakadi:2022lqf,Sarkar:2021ird,Saini:2024mun},
and it is found that the current data is not sufficient to constrain the parameter $\alpha$. 
However, an upper limit on $\alpha$ i.e., $\alpha \le 39.81$ with $95\%$ C. L. is obtained in \cite{Saini:2024mun}
from Planck-2018 and BICEP/Keck array observations by performing a detailed statistical  analysis. A two field inflation model
with both the field having $\alpha$-attractor potentials have also been studied in \cite{Rodrigues:2020fle}.
The $E$-model potential in brane inflation has also been studied in \cite{Sabir:2019wel}. 
$R^2$ attractors and $\alpha$-attractors inflationary models as a subclass of $f(R)$ gravity in the Jordan frame have been 
studied in \cite{Odintsov:2020thl} (see also \cite{Odintsov:2023ypt} for various phenomenological aspects of these models).
In \cite{SantosdaCosta:2020dyl} the variations from Starobinsky potential in the Einstein frame have been studied based on a 
potential derived from brane inflation.  Extensions of Starobinsky model with $R^3$, $R^4$ and
$R^{3/2}$, in addition to $R^2$ term, have also been studied in \cite{Ivanov:2021chn} in the light of Planck 2018 observations. 
It is shown in \cite{Gialamas:2025ofz} that the Starobinsky inflation with an
additional $R^3$ term is in full agreement with recent ACT observations. 

In this work we investigate a potential obtained by the combination of power law and $\alpha$-Starobinsky model in the 
Einstein frame having two parameters $\alpha$ and $\beta$ to analyze the deviation from the Starobinsky inflation.
 Since the required conformal transformations
from Jordan frame to Einstein frame are related to the K\"ahler potential \cite{Cremmer:1978hn,Cremmer:1982en}, we 
consider $R^\beta$ term in the Einstein-Hilbert action and derive the Einstein frame potential by introducing the
parameter $\alpha$ in the conformal transformations.
We employ ModeChord \cite{2021ascl}  an updated version of ModeCode \cite{Mortonson:2010er} to solve the background and 
perturbation equations for the inflaton without assuming the slow-roll approximation, enabling the computation of 
primordial power spectra for scalar and tensor perturbations. These  power spectra
are used in CAMB \cite{Lewis:1999bs} to generate angular power spectra for CMB 
anisotropies and polarization. We then perform MCMC analysis using 
CosmoMC \cite{Lewis:2002ah}  to directly constrain the parameters of 
inflaton potential and number of e-foldings using Planck-2018, BICEP/Keck array and other LSS observations. 
 
We also calculate Bayesian evidence for power law model, $\alpha$-Starobinsky model,  power law $\alpha$-Starobinsky model 
and Starobinsky model using the MCEvidence  \cite{Heavens:2017afc}. which is a tool 
in Python that  computes the Bayesian evidences using 
data from MCMC simulations. It uses a method called $k$-th nearest neighbor to 
estimate how likely each model is, which is 
important when comparing different models in Bayesian analysis. By computing the 
Bayes factors it can be determined that
how well these generalized models align with or outperform the original 
Starobinsky model in  the light of current CMB and LSS observations. 

The rest of this paper is  organized as follows. In section \ref{themodel}, we introduce the  power 
law $\alpha$ Starobinsky model and derive the Einstein frame potential for it from $f(R)$ gravity. Section \ref{theoreframe} provides a comprehensive overview of the background and perturbation
  equations used in ModeChord to compute the primordial power spectra.
We present the results of testing the inflationary models against observational data using Markov chain Monte Carlo (MCMC
) analysis in section \ref{observframe} and in section \ref{baysmodel}, we describe the insights gained through Bayesian 
evidence calculations. Finally, in Section \ref{conclusions}, we summarize our conclusions.

\section{Power law $\alpha$-Starobinsky Model} \label{themodel}
The modified Starobinsky model, which includes an $R^{\beta}$ correction, has been analyzed in
\cite{DeFelice:2010aj,Chakravarty:2014yda,Motohashi:2014tra}. The corresponding action in Jordan frame is given by,
\begin{equation}
S_J = \frac{-M_{Pl}^2}{2}\int \sqrt{-g} f(R) d^4 x,\label{SJ} \quad \quad f(R) = \left(R+\frac{1}{6M^2}\frac{R^{\beta}}{M_{Pl}^{2\beta-2}}\right), 
\end{equation}
where $M_{Pl}^2 = (8\pi G)^{-1}$, $g$ is the determinant of the metric $g_{\mu \nu}$ having signature $+,-,-,-$ and $M$ is a 
real parameter having dimension of mass. We will take  
$M_{Pl}=1$ in the the rest of the paper, and $M$ becomes dimensionless in this unit.  
To obtain the action in the Einstein frame we use conformal transformations 
$\tilde{g}_{\mu\nu}(x) = \Omega(x) g_{\mu\nu}(x)$, where a tilde denotes the 
quantities in the Einstein frame.
The Ricci scalar $R$ defined in the Jordan frame can be expressed in terms of the Einstein frame Ricci scalar $\tilde{R}$ via 
the conformal transformation connecting the two frames as
\begin{equation}
R = \Omega \left(\tilde{R} + 3\widetilde{\Box}\omega - \frac{3}{2}\tilde{g}^{\mu \nu}\partial_{\mu}{\omega}\partial_{\nu}{\omega}\right),\label{ricci}
\end{equation}
here $\omega \equiv \ln\Omega$, $\widetilde{\Box}\omega \equiv \frac{1}{\sqrt{-\tilde{g}}}\partial_{\mu}\left(\sqrt{-\tilde{g}}\tilde{g}^{\mu\nu}\partial_{\nu}{\omega}\right)$ and  $\partial_{\mu}{\omega} = \frac{\partial \omega}{\partial \tilde{x}^{\mu}}$.

To obtain action in the Einstein frame we choose $\Omega = F=\frac{\partial f(R)}{\partial R}$ and also introduce a new scalar 
field $\chi$  defined as
\be
\chi \equiv \frac{\sqrt{6\alpha}}{2}\ln F. \label{fchi} 
\ee
This  leads to 
\be
\Omega = \exp\left(\frac{2\chi}{\sqrt{6\alpha}}\right). \label{omegachi} 
\ee

Here, the parameter $\alpha$ is a dimensionless
parameter, and it is introduced to obtain the potential of $\alpha$-Starobinsky inflation in the Einstein frame. 
Now, using Eq.~\eqref{ricci} and (\ref{omegachi}) along with $\sqrt{-g}=\Omega^{-2}\sqrt{-\tilde{g}}$, the action \eqref{SJ} 
can be transformed to the Einstein-Hilbert form as
\begin{equation}
S_E = \int d^4x\sqrt{-\tilde{g}}\left(-\frac{1}{2}\tilde{R}+\frac{1}{2}\tilde{g}^{\mu \nu}\partial_{\mu}{\chi}\partial_{\nu}{\chi}-V(\chi)\right),\label{SE}
\end{equation}
where the Einstein frame potential $V(\chi)$ is given as
\begin{equation}
V(\chi) = \frac{(RF(R)-f(R))}{2F(R)^2}. \label{potfr}
\end{equation}

Thus, the potential for the power law $\alpha$-Starobinsky model can be expressed as

\begin{equation}
V({\chi}) = (\frac{{\beta}-1}{2})\left(\frac{6M^2}{{\beta}^{\beta}}\right)^{\frac{1}{\beta-1}} \exp\left[{\frac{2\chi}{\sqrt{6\alpha}}\left(\frac{2-\beta}{\beta-1}\right)}\right]\times \left(1-\exp\left(-\frac{2\chi}{\sqrt{6\alpha}}\right)\right)^\frac{\beta}{\beta-1}. \\ \label{alphabetapot}
\end{equation}
We can see that this potential reduces to the Starobinsky ($R^2$) potential in the Einstein frame for $\beta=2$ and $\alpha=1$.
In the large field limit $\chi \gg 2\sqrt{6}$ and $1 < \beta < 2$ the potential from \eqrf{alphabetapot} simplifies to
\be
V({\chi}) = (\frac{{\beta}-1}{2})\left(\frac{6M^2}{{\beta}^{\beta}}\right)^{\frac{1}{\beta-1}} \exp\left[{\frac{2\chi}{\sqrt {6\alpha}}\left(\frac{2-\beta}{\beta-1}\right)}\right].\label{approxpot}
\ee
As there is a correspondence between the $R^2$ (de Sitter) and $R+R^2$ (Starobinsky)
theories of gravity and no-scale supergravity \cite{Ellis:2017xwz}, the supergravity
Lagrangian written in the form of Einstein-Hilbert action requires a conformal
transform expressed in terms of K\"ahler potential $\Omega^2= \exp(-K/3)$
\cite{Cremmer:1978hn,Cremmer:1982en}. As the
potential for $\alpha$-Starbinsky inflation can be obtained by changing the
K\"ahler curvature $\alpha \ne 1$, the corresponding conformal transformation can be
expressed in terms of  K\"ahler potential as $\Omega^2= \exp(-K/(3\alpha))$
\cite{Ellis:2019bmm}. This justifies the introduction of parameter $\alpha$ in \eqrf{fchi}, where the corresponding Jordan 
action, (\ref{SJ}), has $R^\beta$ term. Moreover, for $\alpha=1$, the potential (\ref{approxpot}) reduces to the equivalent 
no-scale supergravity potential obtained of $R^\beta$ model  \cite{Chakravarty:2014yda}.

\begin{figure}[h]
\begin{center}
\includegraphics[width=12cm, height = 7cm]{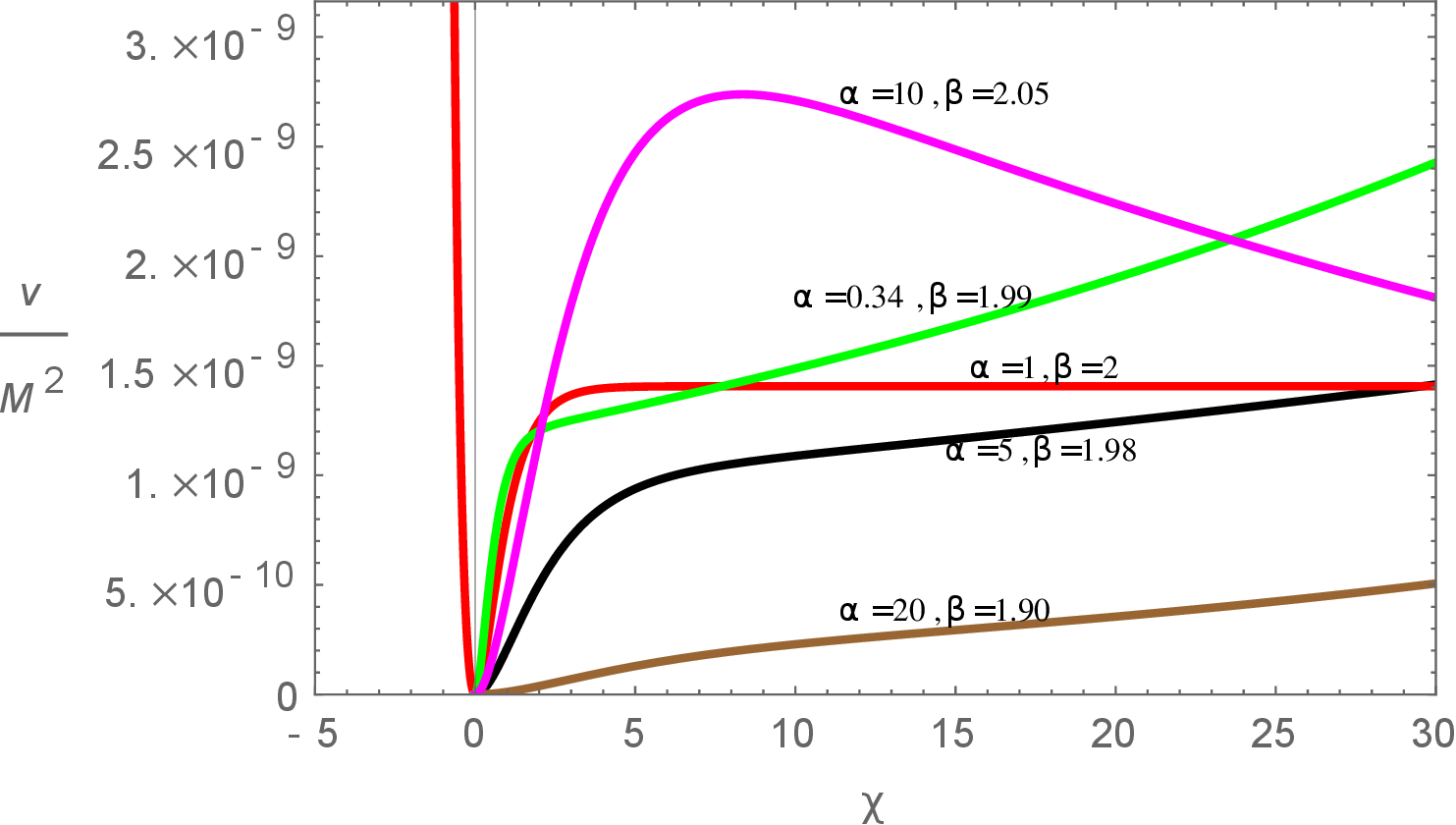}
\caption{Variation of potential (\ref{alphabetapot}) with the field for different values of $\alpha$ and $\beta$, as indicated. Both the potential and the field are expressed in $M_{Pl} = 1$ units.}
\label{fig:alphabetaplot}
\end{center}
\end{figure}

The plot \ref{fig:alphabetaplot} depicts the variation of potential (\ref{alphabetapot}) for different combinations of the parameters $\alpha$ and $\beta$. For $\beta=2$ and $\alpha=1$, the behavior is representative of the Starobinsky $R^2$ inflationary model.
When $\beta$ deviates from 2:
\begin{itemize}
\item{} For $\beta<2$ (e.g.,$\alpha=20$, $\beta=1.90$),the potential becomes steeper. This may correspond to scenarios with larger tensor to scalar ratio in comparison to $\beta=2$.
\item{} For $\beta>2$ (e.g.,$\alpha=10$, $\beta=2.05$), the potential initially rises, reaches a maximum, and then decreases sharply, transitioning away from the plateau into a steeper.
\end{itemize}

The parameter $\alpha$ further modulates the steepness:
\begin{itemize}
\item{} Larger values of $\alpha$ (e.g.,$\alpha=20$) stretch the potential horizontally and amplify the steepness of the potential for a fixed $\beta$, causing potential to grow more rapidly with $\chi$.
\item{} Smaller values of $\alpha$ (e.g.,$\alpha=0.34$) result in more gradual increases in potential. 
\end{itemize}
Overall, the interplay between $\alpha$ and $\beta$ governs the inflationary dynamics. This plot is crucial for exploring how these parameters influence observable predictions.

\section{Inflationary dynamics and power spectra} \label{theoreframe}
In this section we describe the background  equations  for inflationary dynamics and the perturbation equations, which 
are solved numerically using ModeCode. Here, the cosmic time derivatives will be indicated using an overdot notation, 
while differentiation with respect to number of e-foldings ($N$) will be denoted by a prime symbol. 
We will express all equations in natural units i.e. $c=\hbar=8\pi G =1.$
\subsection{ Background equations}
To describe the inflationary dynamics in the Einstein frame, the Einstein-Hilbert action is given as
\be
S_{E} = \int d^4x\sqrt{-\tilde{g}}\left(-\frac{1}{2}\tilde{R}+\frac{1}{2}\tilde{g}^{\mu \nu}\partial_{\mu}{\chi}\partial_{\nu}{\chi}-V(\chi)\right),
\label{se}
\ee
here \(R\) and \(\chi\) are the Ricci scalar and scalar field respectively, and \(V(\chi)\) represents the potential of the 
scalar field.

The line element is given by
\[
ds^2 = dt^2 - a^2 \delta_{ij} dx^i dx^j,
\]
where \(a\) is the scale factor. During inflation the expansion of the Universe is governed by the 
Friedmann equations 
\be
\quad \quad \quad \quad H^2 = \frac{1}{3} \left( \frac{1}{2} \dot{\chi}^2 + V(\chi) \right), \label{H2}
\ee
and
\be
\dot{H} = -\frac{1}{2}  \dot{\chi}^2,  \label{fe1}
\ee

where \(H\) is the Hubble parameter, $H = \frac{\dot a}{a}$.

The  equation of motion for the $\chi$ field in the expanding universe  is

\be
\ddot{\chi} + 3H \dot{\chi} + \frac{dV}{d\chi} = 0. \label{fe2}
\ee

Once we apply the initial conditions, the evolution of universe can be fully specified by any two of \eqrf{H2} to \eqrf{fe2}. 
These equations are collectively called as background equations. Since we use the number of e-foldings, $N=\ln a$, as an 
independent variable for numerical solutions, we can express the background equations in terms of $N$ as follows
\bea
H^2 &=& \frac{\frac{1}{3}V(\chi)}{1-\frac{1}{6}\chi^{\prime 2}},\label{H2n}\\
H^\prime &=& -\frac{1}{2}H\chi^{\prime 2},\label{Hprimen}
\eea
and 
\be
\chi^{\prime \prime}+\left(\frac{H^\prime}{H}+3\right)\chi^\prime+\frac{1}{H^2}\frac{dV(\chi)}{d\chi} = 0.\label{evon}
\ee
where primes denote derivatives taken with respect to the number of e-folds 
$N$. These Background equations can be solved numerically along with perturbation equations.

\subsection{Perturbations Equations}
 To describe primordial scalar perturbations generated during inflation we use  the gauge-invariant comoving curvature 
perturbation $\mathcal{R}$, which is related to the Mukhanov–Sasaki variable $u$ as $u=-z \mathcal{R}$ 
\cite{Mukhanov:1988jd,Sasaki:1986hm}. Here, $z = \frac{1}{H}\frac{d\chi}{d\tau}$ and $\tau$ represents the conformal time. The 
comoving curvature perturbation ${\mathcal {R}}$ remains conserved on super-Hubble scales. Both $u$ and ${\mathcal {R}}$ are
combinations of scalar perturbations in the metric and inflaton perturbations. 

%
The Fourier modes $u_k$ of the Mukhanov-Sasaki variable  obey the equation \cite{Stewart:1993bc,Mukhanov:1985rz,Mukhanov:1990me}, 
\begin{equation}
\frac{d^2 u_k}{d\tau^2} + \left(k^2 - \frac{1}{z}\frac{d^2z}{d\tau^2}\right)u_k = 0. \label{ukeq}
\end{equation}
where $k$ is the modulus of wavenumber  $\vec{k}$ of the Fourier components. 
This second-order differential equation for $u_k$ depends on the background dynamics through the quantity $z$ and its derivatives. 

As discussed earlier, the scalar and tensor power spectra produced during inflation are obtained using ModeCode, which numerically 
integrates the background and perturbation equations in terms of the number of e-folds $N$ as the independent variable, without 
assuming slow-roll, the Mukhanov-Sasaki equation for $u_k$, (\ref{ukeq})  can be rewritten in terms of $N$ as,
\be
u_k^{\prime \prime} + \left(\frac{H^\prime}{H}+1\right)u_k^{\prime}+\Biggl\{\frac{k^2}{a^2H^2}-\left[2-4\frac{H^\prime}{H}\frac{\chi^{\prime \prime}}{\chi^\prime}-2\left(\frac{H^\prime}{H}\right)^2 -5\frac{H^\prime}{H}-\frac{1}{H^2}\frac{d^2V}{d\chi^2}\right] \Biggr\}u_k = 0, \label{ukn}
\ee
Similarly the equation for the Fourier components of the tensor perturbations $v_k$ is
\be
\frac{ d^2 v_k}{d\tau^2} + \left(k^2 - \frac{1}{a}\frac{d^2 a}{d\tau^2}\right)v_k = 0,\label{vk}
\ee
which in terms of $N$ becomes
\be
v_k^{\prime \prime} + \left(\frac{H^\prime}{H}+1\right)v_k^{\prime} + \left[\frac{k^2}{a^2H^2} -\left(\frac{H^\prime}{H}+2\right) \right]v_k = 0.\label{vkn}
\ee
\subsection{  Power spectra}
We define the primordial power spectrum of curvature perturbations, $P_{\mathcal{R}}(k)$, in terms of the vacuum expectation 
value of $\mathcal{R}$ as.
\be
\mathcal{P_\mathcal{R}} = \frac{k^3}{2\pi^2}\langle \mathcal{R}_{k}\mathcal{R}_{k^\prime}^*\rangle\delta^3(k-k^\prime), 
\label{Prs}
\ee
which in terms of  $u_k$ is
\be
\mathcal{P_\mathcal{R}}(k) = \frac{k^3}{2\pi^2}\bigg|\frac{u_k}{z}\bigg|^2. \label{Ps}
\ee
Similarly,  the primordial tensor power spectrum can be expressed in terms of   $v_k$ and $z$ as 
\be
\mathcal{P}_{t}(k) = \frac{4k^3}{\pi^2}\bigg|\frac{v_k}{a}\bigg|^2.\label{Pt}
\ee
In ModeCode the scalar and tensor power spectra are obtained by solving mode Eqns.~ (\ref{ukn}) and (\ref{vkn}) numerically along 
with the background Eqns. (\ref{Hprimen}) and (\ref{evon}) using Bunch-Davious initial conditions. 
The scalar spectral index $n_s$ and the tensor spectral index $n_t$ are extracted from the numerically obtained power 
spectra by using their definition as \cite{Bassett:2005xm}.
\be
\quad \quad  n_s = 1 + \frac{d \ln\mathcal{P_\mathcal{R}}}{d \ln k},\label{ns}
\ee
and
\be
n_t = \frac{d \ln\mathcal{P}_{t}}{d \ln k}.\label{nt}
\ee
Similarly, for the tensor-to-scalar ratio $r$,  we use its definition \cite{Bassett:2005xm}
\be
r = \frac{\mathcal{P}_{t}}{\mathcal{P_\mathcal{R}}}.\label{r}
\ee
In our analysis both $n_s$ and $r$ are derived parameters, and the parameters of the potential (\ref{alphabetapot})  
$M$, $\alpha$ and $\beta$ are directly constrained from CMB observations.

\section{Observational constraints} \label{observframe}
To compute the power spectra for scalar perturbations, (\ref{Ps}), and tensor 
perturbations, (\ref{Pt}) we modify publicly available code ModeChord, an updated 
version of ModeCode \cite{Mortonson:2010er} for the power law $\alpha$-Starobinsky 
potential (\ref{alphabetapot}). ModeCode solves background  and perturbations equations, described in 
the previous section, numerically without using slow-roll approximation for a given 
inflaton potential.  We also vary $N_{pivot}$, the number of e-foldings from the end of inflation the the time when the length
scales corresponding to Fourier mode $k_{pivot}$ leave the inflationary horizon, along with the other potential parameters
to take into account the general reheating scenario. The primordial power spectra calculated by ModeCode are used in 
CAMB \cite{Lewis:1999bs}
to compute the angular power spectra for CMB temperature anisotropy and polarization
 for a given set of cosmological parameters. These 
theoretically computed angular power spectra are used in CosmoMC \cite{Lewis:2002ah},
 which performs Markov chain Monte-Carlo analysis to put constraints on the 
parameters of inflaton potential, $N_{pivot}$ and other parameters of $\Lambda$CDM 
model from various CMB and LSS observations. We have used Planck-2018 TT, TE, EE, lowE with lensing, 
BICEP (BK18) \cite{BICEP:2021xfz}, Dark Energy Survey \cite{DES:2020mlx}, and BAO data from BOSS/DR12 \cite{BOSS:2015zan}, 
 6dFGS and SDSS  to put constraints on the parameters $M$, $\alpha$,  $\beta$ of inflaton potential, $N_{pivot}$ and 
other $\Lambda$CDM parameters. CosmoMC uses Bayesian statistics to compute the posterior probabilities for various parameters, which
requires the prior probabilities. We use flat priors for the potential parameters and $N_{pivot}$ given in 
Table \ref{Tab:prior}, while the priors for other $\Lambda$CDM parameters are used as in \cite{Planck:2018vyg}. The priors for $M$ 
and $\beta$ are based on earlier analysis of power law Starobinsky inflation 
\cite{Motohashi:2014tra,Odintsov:2022bpg,Meza:2021xuq,Saini:2023uqc}. 
As the parameter $\alpha$ for the $\alpha-$Starobinsky inflation is not well constrained by recent 
observations \cite{Saini:2024mun} and there is no theoretical estimate for it, we have chosen a broad range of the priors for
$\alpha$ to obtain a better estimates for it and its relation with other parameters.   For each parameter the MCMC convergence diagnostic tests is performed over
the four chains using the Gelman and Rubin variance of mean/mean of chain variance 
R-1 statistics. The marginalized joint probability distributions are plotted 
using  GetDist tool \cite{Lewis:2019xzd}.    

\begin{table}[h]
\centering
\resizebox{0.3\textwidth}{!}{%
\begin{tabular}{|c|c|}
\hline
\textbf{Parameter} & {\textbf{Prior range}} \\ 

\hline
$\mathbf{N_{\text{pivot}}}$ & [25 : 90]  \\
\hline
$\mathbf{\log_{10} M}$ & [-8.0 : -3.0]   \\
\hline
$\mathbf{\beta}$   & [1.90 : 2.07]     \\
\hline
$\mathbf{\log_{10} \alpha}$ & [-8.0 : 4.0]  \\
\hline
\end{tabular}
}
\caption{Priors on $N_{\text{pivot}}$ and model parameters for power law $\alpha$-Starobinsky model.}
\label{Tab:prior}
\end{table}

Table \ref{Tab:constraint}  shows  the observational constraints on 
the parameters of power law $\alpha$-Starobinsky potential (\ref{alphabetapot}) and $N_{pivot}$, obtained 
from the Planck 2018 observations  in combination with BAO and BICEP 
(BK18) \cite{BICEP:2021xfz}.

\begin{table}[h]
\centering
\begin{tabular}{| c| c |c| c |}
\hline
 Parameter &  68\% limits & 95\% limits   &  99\% limits  \\
\hline

{\boldmath$N_{\text{pivot}}$} & $47^{+9}_{-4}$ & $47\pm{10}$ & $47^{+10}_{-20}$ \\
\hline

{\boldmath$\log_{10} M$} & $-4.45^{+0.16}_{-0.12}$ & $-4.45^{+0.24}_{-0.29} $ 
& $-4.45^{+0.32}_{-0.34}$\\
\hline

{\boldmath$\beta$} & $1.969^{+0.018}_{-0.0086}$ & $1.969^{+0.020}_{-0.023}$ & $1.969^{+0.022}_{-0.027}$ \\
\hline

{\boldmath$\log_{10} \alpha$} & $0.37\pm{0.43}$ & $0.37^{+0.82}_{-0.85} $ 
& $0.4^{+1.0}_{-1.0}$\\
\hline

{\boldmath$n_s$} & $0.9685\pm{0.0031}$ & $0.9685^{+0.0064}_{-0.0060} $ & $0.9685^{+0.0084}_{-0.0081}$ \\
\hline

{\boldmath$r$} & $0.0185^{+0.0060}_{-0.0087}$ & $0.019^{+0.015}_{-0.014} $ 
& $0.019^{+0.024}_{-0.016}$\\
\hline
\end{tabular}
\caption{Constraints on parameters of potential, $r$ and $n_{s}$ using Planck-2018 TT, TE, EE, lowE with lensing,
BICEP (BK18) \cite{BICEP:2021xfz}, Dark Energy Survey \cite{DES:2020mlx}, and BAO data from BOSS/DR12 \cite{BOSS:2015zan},
 6dFGS and SDSS.}
\label{Tab:constraint}
\end{table}

It can be seen from the table that the constraints on the parameters $\beta$ and
$\alpha$ are
\bea
\beta = 1.969^{+0.020}_{-0.023},\, \, \, 95\%\, C.\, L., \label{betadata} \\
\log_{10} \alpha = 0.37^{+0.82}_{-0.85},\, \, \, 95\%\, C.\, L. \label{alphadata}
\eea

This indicates that the current observations allow the deviation from $\beta=2$ by
$2\sigma$, while  $\alpha=1$ is still consistent the observations. The number of
e-foldings obtained for this model is
\be
N_{pivot} = 47\pm 10\, \, \, 95\%\text{C. L.} \label{npivotob}
\ee
\begin{figure}[h]
\begin{center}
\subfigure[]{
 \includegraphics[width=8cm, height=7cm]{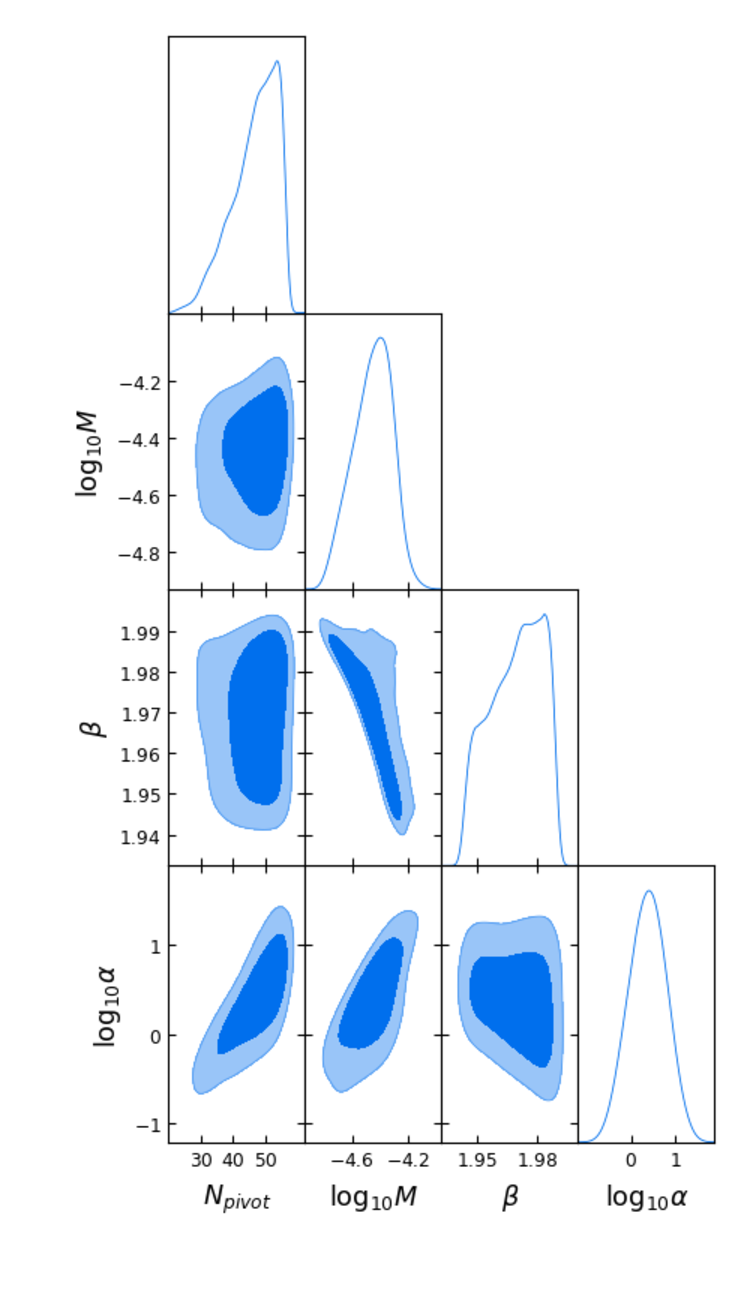}
 \label{fig:modelca}
}
\subfigure[]{
\includegraphics[width=7cm, height=7cm]{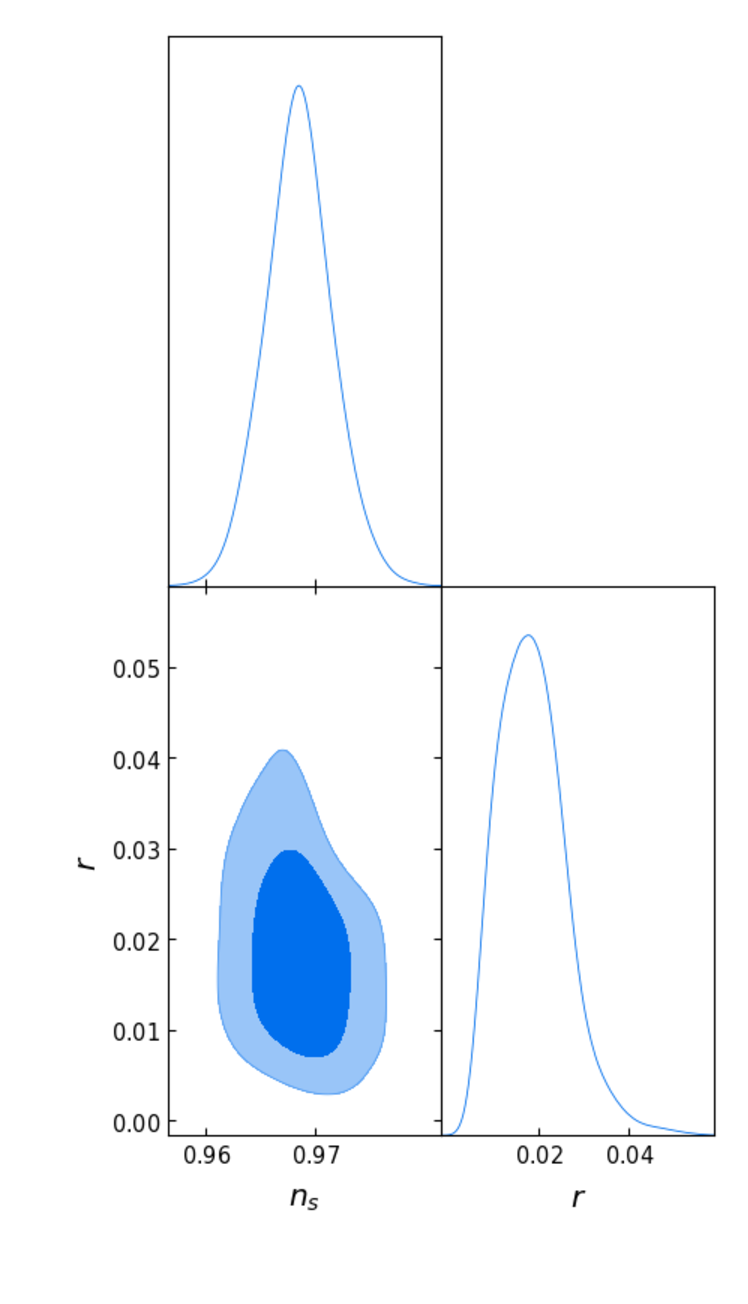}
 \label{fig:modelcb}
}
\caption{1$\sigma$ and 2$\sigma$ confidence contours of (a) potential parameters and $N_{pivot}$, and (b) derived parameters $n_s$ and $r$ using Planck-2018 TT, TE, EE, lowE with lensing,
BICEP (BK18) \cite{BICEP:2021xfz}, Dark Energy Survey \cite{DES:2020mlx}, and BAO data from BOSS/DR12 \cite{BOSS:2015zan},
 6dFGS and SDSS. Marginalized probability distributions of the individual parameters are also displayed for power law $\alpha$-Starobinsky model.}
\label{fig:modelccon}
\end{center}
\end{figure}

Fig.~\ref{fig:modelccon} illustrates the  posterior probability distribution and 
the confidence regions at $68\%$ and $95\%$ confidence levels for the potential 
parameter and the derived parameters $(n_s,r)$.
From the Fig.~\ref{fig:modelca}, it can be seen that the parameters 
$(N_{pivot},\, \log_{10}\alpha)$, $(\log_{10}M, \, \log_{10}\alpha)$ and 
$(\log_{10}M,\,\beta)$ are strongly correlated, 
while $(\log_{10}M,\, N_{pivot})$, $(\beta, N_{pivot})$ are weekly correlated or 
almost uncorrelated. The contours of $(\beta, \log_{10}M)$ are elongated and tilted downward, indicating a negative correlation. As $\log_{10}M$ increases, $\beta$ decreases slightly. For the parameters $(\log_{10}\alpha$, $\log_{10}M)$ contours are stretched and tilted upward, showing a positive correlation. However, in case of 
power law Starobinsky inflation $(\beta,\, N_{pivot})$ and $(\log_{10}M, N_{pivot})$ 
are strongly correlated \cite{Saini:2023uqc},
and in case of $\alpha$-Starobinsky inflation $N_{pivot}$ and $\alpha$ are not
correlated \cite{Saini:2024mun}. 
Fig.~\ref{fig:modelcb} shows the joint $68\%$ and $95\%$ C.L. constraints on scalar spectral index $n_s$ and 
tensor-to-scalar ratio $r$. These are derived parameters, and their constraints are determined by using Eqns. 
(\ref{ns}) and (\ref{r}). As depicted in Fig.~\ref{fig:modelcb}, the derived parameters $n_s$ and $r$ are 
almost uncorrelated, contrary to standard inflationary scenario and 
$\alpha$-Starobinsky inflation \cite{Saini:2024mun}, but, similar to power law 
Starobinsky inflation \cite{Saini:2023uqc}.

\begin{figure}[h]
\begin{center}
\includegraphics[width=8cm, height = 7cm]{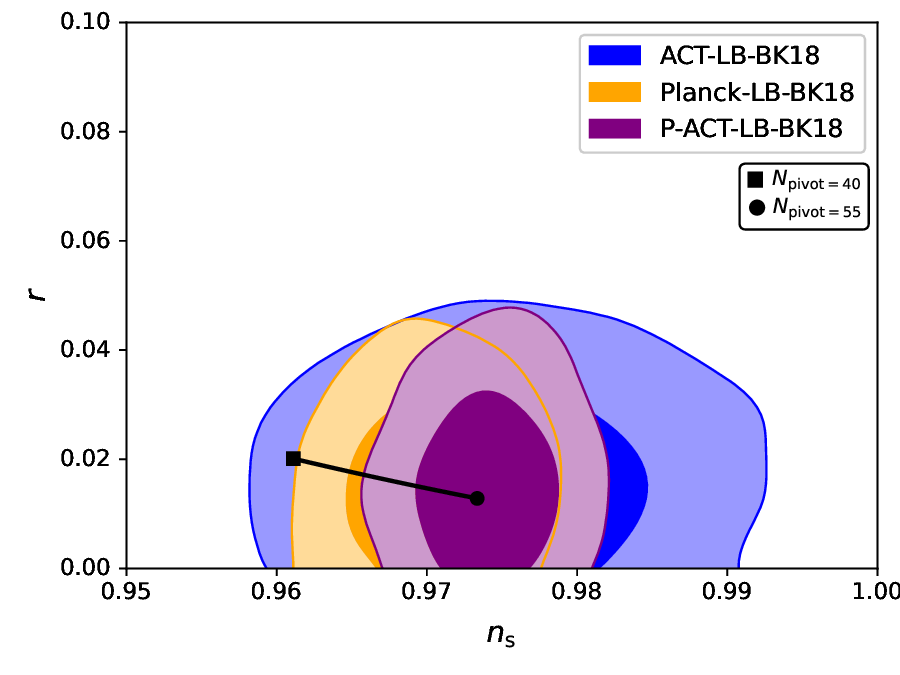}
\caption{The $r-n_s$ predictions  for power law $\alpha$-Starobinsky inflation
 (black line) along with the joint constraints from  ACT, Planck,
BICEP/Keck and BAO observations \cite{ACT:2025tim}.}
\label{fig:actrns}
\end{center}
\end{figure}

Fig.~\ref{fig:actrns} shows the $r-n_s$ predictions for our model with mean values of the potential parameters along with
joint $68\%$ and $95\%$ C.L. constraints from ACT, Planck, BAO from DESI and BICEP
observations \cite{ACT:2025fju,ACT:2025tim}. We choose the
parameters $\log_{10}\alpha= 0.37$, $\beta = 1.969$, $M= 10^{-5}$, and vary number of e-foldings $N_{pivot}$ between $40$ to $55$
to determine the dependence of tensor-to-scalar ratio $r$ on $n_s$ shown by black line in the figure.
 It is clear from Fig.~\ref{fig:actrns} that the $r-n_s$ predictions of our model  agree within $1\sigma$ with the recent ACT
observations.

\section{BAYESIAN MODEL SELECTION} \label{baysmodel}
To determine whether the current observations allow deviations from Starobinsky 
inflation we compute the Bayesian evidences for Starobinsky inflation, power law
Starobinsky inflation, $\alpha$-Starobinsky inflation and power law 
$\alpha$-Starobinsky inflation proposed in this work. 
In problems of statistical inference in cosmology Bayes' theorem combines the prior 
information $\pi(\theta|{\cal M})$ on the parameters $\theta$ of a physical model 
${\cal M}$ with the likelihood $\mathcal L(y|\theta, {\cal M})$, which represents 
the distribution of the data points $y$ for a given parameter choice $\theta$, 
to yield the posterior distribution $p(\theta|y, {\cal M})$,
\be
p(\theta|y, {\cal M}) = \frac{\mathcal L(y|\theta,{\cal  M})
\pi(\theta|{\cal M})}{p(y|{\cal M})}. \label{priorpost}
\ee
From above equation, the evidence is obtained by integrating the unnormalized posterior distribution over the parameter space.
\be
p(y|{\cal M}) = \int d^n\theta \, \mathcal L(y|\theta, {\cal M}) 
\pi(\theta|{\cal M}). \label{evidence}
\ee
In Bayesian model selection, the posterior probability $p(y|{\cal M})$, which 
supports a model ${\cal M}$ based on the data $y$, is required to be computed. 
Bayes' theorem \cite{Trotta:2008qt} allows the conditional probability 
\( p(y|{\cal M}) \) to be inverted, providing \( p({\cal M}|y) \), which is 
expressed as

\begin{equation}
    p({\cal M}|y) = \frac{p(y|{\cal M}) \pi({\cal M})}{p(y)},
\end{equation}

where \( \pi({\cal M}) \) is the prior probability of the model \( {\cal M} \), 
\( p(y|{\cal M}) \) is the likelihood of the data \( y \) given the 
model \( {\cal M} \), and \( p(y) \) is the marginal likelihood or evidence. 
The normalization is given by the model evidence \( p(y) \), which arises from 
summing over the discrete set of models \( {\cal M}_i \). 
Each model \( {\cal M}_i \) has a corresponding prior probability 
\( \pi({\cal M}_i) \) and a Bayesian evidence \( p(y|{\cal M}_i) \), as shown in 
\eqrf{evidence}.
\begin{equation}
    p(y) = \sum_i p(y|{\cal M}_i) \pi({\cal M}_i).
\end{equation}
Typically, the comparison between two or more models is made by calculating the 
ratio of their evidences. The Bayes' ratio
\begin{equation}
    \ln B_{ij} = \ln \left( \frac{p(y|{\cal M}_i)}{p(y|{\cal M}_j)} \right),
\end{equation}
is subsequently interpreted according to Jeffreys' guidelines 
\cite{Jeffreys:1939xee}. The observational support for the underlying model 
${\cal M}_i$ relative to ${\cal M}_j$ is quantified based on the values of 
$\ln B_{ij}$(or alternatively $\ln B_{ij}$). Additionally, according to the 
"revised Jeffreys' scale" by Kass and Raftery \cite{Kass:1995loi}, the values 
of \( |\ln B_{ij}| \) in the intervals \((0, 1.0)\), \((1.0, 3.0)\), \((3.0, 5.0)\), 
and \((5.0, \infty)\) indicate weak, definite/positive, strong and Very strong 
evidence, respectively. 

Here we use the publicly available code MCEvidence \cite{Heavens:2017afc} to 
calculate the Bayesian evidences for the models.  MCEvidence  uses only the 
MCMC chains produced by CosmoMC and provides the logarithm of the Bayes factor, 
$\ln B_{ij}$. This value is then used to evaluate whether model $M_i$ is preferred 
over model $M_j$.

\begin{table}[h]
\centering
\resizebox{0.4\textwidth}{!}{%
\begin{tabular}{|c|c|c|c|}
\hline
 \textbf{Model} &   \textbf{\boldmath{$\ln B$}}     &   \textbf{\boldmath{$\Delta\ln B_{ij}$}}     \\
\hline
 Starobinsky  & -1958.506   &  -     \\
\hline
Power law   & -1957.047  & 1.459 \\
\hline
$\alpha$-Starobinsky  & -1957.054  & 1.452 \\
\hline
 Power law $\alpha$-Starobinsky  & -1956.820   & 1.686 \\
\hline

\end{tabular}
}
\caption{The column $\Delta\ln B_{ij}$ corresponds to the comparison of Model Power law, Model $\alpha$-Starobinsky  and Power law $\alpha$-Starobinsky with Starobinsky Model.}
\label{Tab:evidence}
\end{table}
 
The Bayesian evidences and Bayes factors for various models are shown in 
Table.~\ref{Tab:evidence}. We have computed Bayes factor by treating Starobinsky
inflation as our reference. It can be seen from the Table that the power law 
$\alpha$-Starobinsky inflation has a higher Bayes factor compared to other models, 
hence it is mildly favored by current observations.

\section{Conclusions} \label{conclusions}
One of the best suited model from Planck 2018 \cite{Planck:2018jri} and 
BICEP/Keck (Bk18) observations \cite{BICEP:2021xfz} is Starobinsky inflation
\cite{Starobinsky:1980te}, where the inflation can be achieved without scalar field
by adding $R^2$ term in the Einstein-Hilbert action. An interesting feature of 
Starobinsky inflation is that its potential in the Einstein frame can be realized 
in the framework of no-scale supergravity with a non-compact 
$SU(2,1)/SU(2)\times U(1)$ symmetry \cite{Ellis:2013xoa}, where we have a modulus 
field fixed by the other dynamics and the inflaton field is a part of the chiral 
superfield with a minimal Wess-Zumino superpotential. 
There are generalizations of Starobinsky inflation such as power law, $R^\beta$, 
Starobinsky inflation  based on higher order metric theories of gravity 
\cite{Schmidt:1989zz,Maeda:1988ab,Muller:1989rp,Gottlober:1992rg,DeFelice:2010aj,Nojiri:2010wj,Nojiri:2017ncd}, and $\alpha$-Starobinsky inflation ($E$-model) based on
supergravity \cite{Ellis:2013nxa,Ferrara:2013rsa,Kallosh:2013yoa}, where the 
inflaton potential has a parameter $\alpha$ in the exponential in the Einstein frame.
The Einstein frame potential for the power law Starobinsky inflation can also be 
derived from no-scale supergravity \cite{Chakravarty:2014yda}.

In this work we examine a potential formulated by combining the power law and 
$\alpha$-Starobinsky model, \eqrf{alphabetapot}, in the Einstein frame, which contains
both the parameters $\alpha$ and $\beta$ to analyze the deviation from Starobinsky
inflation in the light of current observation. We obtain the potential for 
power law $\alpha$-Starobinsky inflation by considering $R^\beta$ term in the 
Einstein action and introducing a parameter $\alpha$ in the conformal transformation
via field redefinition motivated by supergravity 
\cite{Cremmer:1978hn,Cremmer:1982en}. 
We compute the primordial power spectra numerically using ModeChord
\cite{2021ascl} by solving background and perturbation equations without using the 
slow-roll approximation in the Einstein frame. With the help of this we perform
MCMC analysis using CosmoMC to put constraints on the parameters $M$, $\alpha$
and $\beta$ of the potential (\ref{alphabetapot}) and the number of e-foldings
$N_{pivot}$ from the end of inflation to the time when the scales corresponding to
$k_{pivot}$ left the Hubble radius during inflation. We find 
$\log_{10} \alpha = 0.37^{+0.82}_{-0.85}$ and 
$\beta = 1.969^{+0.020}_{-0.023}$ at  $95\%$ C.L. This indicates that the current 
CMB and LSS observations allow $2\sigma$ deviation from $\beta=2$. However, the limits
on $\log_{10}\alpha$ are larger than the central value indicating that  $\alpha=1$
is consistent with the current observations. These results are similar to power law
Starobinsky inflation \cite{Saini:2023uqc} and $\alpha$-Starobinsky 
inflation  \cite{Saini:2024mun}. 
We also find the parameter that determines the energy scale of inflation 
$M= \left(M=3.34^{+2.62}_{-1.73}\right)\times 10^{-5}$ at $95\%$ C. L, and the 
number of e-folds from the end of inflation to the time when the pivot scale 
$k_{pivot}$ leaves the inflationary horizon $N_{pivot} = 47\pm 10$ at $95\%$ C. L.
Interestingly, by performing MCMC analysis, we find that the parameters $\beta$ and
$N_{pivot}$ are uncorrelated, which is in contrast to power law Starobinsky inflation
\cite{Saini:2023uqc} where these parameters are strongly correlated. 
Similarly, the parameters $\log_{10}\alpha$ and $N_{pivot}$ are strongly correlated 
in contrast to $\alpha$-Starobinsky inflation \cite{Saini:2024mun} where these are 
uncorrelated. The parameters $\log_{10}M$ and $N_{pivot}$ are weakly correlated 
similar to $\alpha$-Starobinsky inflation \cite{Saini:2024mun}, but, contrary 
to power law Starobinsky inflation \cite{Saini:2023uqc}. There is also no
correlation between the derived parameters $n_s$ and $r$ contrary to 
$\alpha$-Starobinsky inflation \cite{Saini:2024mun}, but, similar to power law 
Starobinsky inflation \cite{Saini:2023uqc}. 

Recently the Atacama
Cosmology Telescope, in combination with Planck 2018, BAO data from DESI Y1
\cite{ACT:2025fju}, has reported $n_s=0.9743\pm 0.0034$, which is slightly higher than the value reported by  Planck 2018
observations, i.e., $n_s=0.9651\pm 0.004$ \cite{Planck:2018jri}.  This suggests that the Starobinsky model is now disfavored at the $2\sigma$ level
\cite{ACT:2025tim} as it predicts lower values for spectral index. We also
obtain $r-n_s$ predictions for our model for the mean values of the potential
parameters $\alpha$ and $\beta$ and varying $N_{pivot}$ between $40$ to $55$. We
find that these predictions lie with $68\%$ C.L. of joint constraints from
ACT, Planck-2018, BICEP and BAO data from DESI Y1 \cite{ACT:2025fju,ACT:2025tim}.

To determine which of $R^{\beta}$, $\alpha$-Starobinsky, or power law 
$\alpha$-Starobinsky model is best suited by the observation,  we  calculate the 
Bayesian evidences of each model using MCEvidence \cite{Heavens:2017afc}. 
We also calculate  the logarithm of the Bayes factor, $\ln B_{ij}$ given 
in \eqrf{evidence} for each of the three models. Here, $i$ represents $R^{\beta}$, $
\alpha$-Starobinsky, and power law $\alpha$-Starobinsky model, while $j$ represents 
the Starobinsky model as reference model.  So we can conclude from table \ref{Tab:evidence} that power law $\alpha$-Starobinsky model has higher Bayesian evidence as a result it shows greater deviation from Starobinsky model as compared to model $R^{\beta}$ and model $\alpha$-Starobinsky. Hence, power law $\alpha$-Starobinsky model is mildly favored over the other two models.
 
Generalizations of Starobinsky inflation have both theoretical and phenomenological implications.  
As the original Starobinsky model is now disfavored at the $2\sigma$ level by the recent ACT observations
\cite{ACT:2025tim}, it highlights the need to study various generalizations of Starobinsky inflation. 
Moreover, the potential for various generalizations of Starobinsky inflation can be incorporated in the framework of supergravity, which can have several implications
for particle physics phenomenology \cite{Ellis:2016ipm,Ellis:2017jcp,Ellis:2019jha,Ellis:2019opr}.

\end{document}